\documentstyle[12pt,epsf]{article}
\newcommand{\re}{\par\hangindent=0.5cm\hangafter=1\noindent}

\newcommand{\lsim}{\raisebox{0.3mm}{\em $\, <$} \hspace{-3.3mm}
\raisebox{-1.8mm}{\em $\sim \,$}}
\newcommand{\gsim}{\raisebox{0.3mm}{\em $\, >$} \hspace{-3.3mm}
\raisebox{-1.8mm}{\em $\sim \,$}}
\newcommand{\bm}[1]{\mbox{\boldmath $#1$}}
\begin{document}

\baselineskip 7mm

\begin{center}
{\Large\bf Radiatively Driven Mass Accretion onto Galactic Nuclei
by Circumnuclear Starbursts}\\[5mm]
Masayuki Umemura\\
{\it Center for Computational Physics, University of Tsukuba, Tsukuba, 
Ibaraki 305, Japan; umemura@rccp.tsukuba.ac.jp}\\
Jun Fukue\\
{\it Astronomical Institute, Osaka Kyoiku University, 
Asahigaoka, Kashiwara, Osaka 582, Japan; fukue@cc.osaka-kyoiku.ac.jp}\\
and
Shin Mineshige\\
{\it Department of Astronomy, Kyoto University, Sakyo-ku, 
Kyoto 606, Japan; minesige@kusastro.kyoto-u.ac.jp}\\[1cm]
{\it To appear in M.N.R.A.S.}
\end{center}

\newpage

\baselineskip 5.4mm

\begin{center}
{\large\bf ABSTRACT}
\end{center}

We examine the physical processes regarding
the radiatively driven mass accretion onto galactic nuclei due to 
the intensive radiation from circumnuclear starbursts.
The radiation from a starburst not only contracts
an inner gas disk by the radiation flux force, 
but also extracts angular momenta due to
the relativistic radiation drag, thereby inducing an avalanche of the
surface layer of disk. To analyze such a mechanism, 
the radiation-hydrodynamical
equations are solved with including the effects of 
radiation drag force as well as radiation flux force.
As a result, it is found that
the mass accretion rate due to the radiative avalanche is given by
$\dot{M}(r)= \eta (L_\ast / c^2)(r/R)^2 (\Delta R/R)(1-{\rm e}^{-\tau}) $
at radius $r$, where the efficiency $\eta$ ranges from 0.2 up to 1, 
$L_\ast$ and $R$ are respectively
the bolometric luminosity and the radius of a starburst ring, 
$\Delta R$ is the extension of the emission regions, and $\tau$ is
the face-on optical depth of a disk.
In an optically thick regime, 
the rate depends upon neither the optical depth nor
the surface mass density distribution of the disk.
The present radiatively driven mass accretion may 
provide a physical mechanism which enables
mass accretion from 100 pc scales down to $\sim$pc, and
it can eventually link to an advection-dominated viscous accretion
onto a massive black hole.
The radiation-hydrodynamical and self-gravitational instabilities 
of the disk are briefly discussed. In particular,
the radiative acceleration possibly builds up a dusty wall,
which shades the nucleus in edge-on views. 
It provides another version for the formation of an obscuring torus.

\vspace{3mm}

\newpage

\section{Introduction}

In a standard picture of active galactic nuclei (AGNs), a viscous accretion disk 
of $\lsim$pc surrounding a putative supermassive black hole
is thought to be the central engine for the enormous energy output.
The mass supply onto such tiny regions is an issue of growing interest.
The accretion driven by a non-axisymmetric wave or a bar
is a possible mechanism
(Shore \& White 1982; Shlosman, Frank, \& Begelman 1989),
but it is found so far that the bar-driven
accretion is just effective in the  
galactic nuclear regions beyond $\sim$100 pc (Wada \& Habe 1995).
Hence, another mechanism that enables the mass accretion
from 100 pc down to pc scales is required.
Recently, Umemura, Fukue, \& Mineshige (1997) have proposed a possible 
physical mechanism in the relevant scales, that is, the mass accretion
driven by radiation from circumnuclear starbursts.

In recent years, there have been accumulated observational evidences for starbursts 
in circumnuclear regions of AGNs. 
Among them, it is remarkable that molecular gas of at least 
$10^{11} M_\odot$ and dust of $\sim 10^{8} M_\odot$
are found towards a QSO (BR1202-0725) at redshift $z=4.69$
(Ohta et al. 1996; Omont et al. 1996).
This, taking into account the Universe age of
$\sim 1$Gyr at $z=4.69$ and the inferred metallicity, suggests that
the QSO inhabits a massive host galaxy undergoing starbursts.
Also, the recent {\it HST} images of nearby luminous QSOs have
shown that luminous QSO phenomena occur preferentially in
luminous host galaxies, often being ellipticals 
(McLeod \& Rieke 1995; Bahcall et al. 1997; Hooper, 
Impey, \& Foltz 1997).  
In addition, AGN events appear to be frequently accompanied by starbursts 
(Soifer et al. 1986; Scoville et al. 1986; Heckman 1991; Rieke 1992; 
Scoville 1992; Fillipenko 1992).  
The {\it IRAS} survey has revealed that
at bolometric luminosities greater than $10^{12}L_\odot$, the
space density of starburst galaxies are just comparable to that
of QSOs in the same luminosity range (Soifer et al. 1986).
Millimeter observations of the IR galaxies indicate
that all are exceedingly rich in molecular gas with up to
$2\times 10^{10}M_\odot$ in their central regions (Scoville et al. 1986).
The correlations between X-ray and CO luminosities as well as
between CO and far-infrared luminosities
in Seyfert galaxies also suggest that more powerful AGNs inhabit
more active star-forming galaxies (Yamada 1994).
Such increasing evidences suggest a possible physical connection
between AGN events and starburst activities 
(Norman \& Scoville 1988; Rowan-Robinson \& Crawford 1989; 
Terlevich 1992; Scoville 1992; Terlevich 1992;
Perry 1992; Taniguchi 1992; Rowan-Robinson 1995).

As for the structures of starburst regions,
recent high-resolution observations of Seyfert nuclei, 
including those from {\it HST},
have revealed circumnuclear starburst regions
with radial extension of a few tens of pc up to kpc, which often exhibit 
ring-like features  
(Wilson et al. 1991; Forbes et al. 1994; Mauder et al. 1994;
Buta, Purcell, \& Crocker 1995; Barth et al. 1995; Maoz et al. 1996;
Storchi-Bergman, Wilson \& Baldwin 1996; Leitherer et al. 1996).
The rings frequently contain a number of extremely compact star clusters
of $<5$ pc, which are composed of hundreds of O-type stars.
Recently, an unexpectedly large ($\sim 100$ pc) disk of gas and 
dust is discovered by the {\it HST} 
around a bright nucleus in an active galaxy NGC 4261
(Jaffe et al 1993; Ferrarese, Ford, \& Jaffe 1996). 
This disk is surrounded by a fat torus composed of early-type stars. 

Umemura, Fukue, \& Mineshige 
(1997, hereafter referred to as Paper I) 
have considered the radiative effects on nuclear disks due to
circumnuclear starbursts, and proposed the possibility 
of a radiative avalanche, that is, the mass accretion 
driven by the relativistic radiation drag.
In paper I, the effects of radiation drag are analyzed 
solely by solving the azimuthal equation of motion.
In this paper, we explore extensively 
the radiative avalanche with including radial motion driven by
radiation flux force. 
Also, the radiation-hydrodynamical instability of the nuclear region
is analyzed with including vertical motion as well, so that the 
radiative building-up of a dusty obscuring wall is predicted.
Some related topics such as the supernova effects, the generation of
seed magnetic fields, and a possible link to an advection-dominated
viscous flow are discussed.
In \S 2, the basic equations which
govern the radiation-hydrodynamical disk evolution are provided. 
In \S 3, the timescales for a radiative avalanche, the disk evolution,
and the resultant mass accretion rate are presented. In \S 4, 
the radiation-hydrodynamical/gravitational instabilities are analyzed.
In \S 5, we discuss the effects of stellar evolution, the generation
of magnetic fields, and a link between a radiative avalanche
and an advection-dominated accretion onto a black hole. 
\S 5 is devoted to the conclusions.

\section{Basic Equations}

\subsection{Radiation Fields by Starbursts}

Here, we suppose a geometrically thin ring of starburst regions
for simplicity, so that the assumption allows us to 
explore the radiative effects
analytically. The effects of the extension of the starburst regions
are briefly discussed. Under the assumption, the radiation fields by 
the starburst ring with the radius $R$ and the bolometric luminosity $L_\ast$
are given as follows (see also Paper I):
The radiation energy density at $r(<R)$ is given by
\begin{equation}
	E={ L_\ast \over 4 \pi c (R^2-r^2) }. \label{rade}
\end{equation}
The radiation flux in radial directions is given by
\begin{equation}
	F^r= {L_\ast \over 8\pi R^2}
{2x(x^2+1)G_1-3xG_2 \over (x^2+1)^{5/2}} ,
\end{equation}
where $x \equiv r/R$,  and $G_1$ are $G_2$ are respectively
$
	G_1=G[3/4, 5/4, 1, 4x^2/(x^2+1)^2],
$
and
$
	G_2=G[5/4, 7/4, 2, 4x^2/(x^2+1)^2]
$
with $G$ being the Gaussian hypergeometric function.
This solution is well approximated by
\begin{equation}
	F^r \simeq -{L_\ast \over 4\pi R^2}\left[ {r \over 2R} +
	{r^3 \over \pi R^2 (R-r)} \right]. \label{radfr}
\end{equation}
[This is slightly different from equation (2) in Paper I, but
this is a more accurate approximation.]
The rotation of the ring generates the toroidal flux at $r$ to be 
\begin{equation}
	F^\varphi={ 3E Vr \over 2R }, \label{radfp}
\end{equation}
where $V$ is the rotation velocity of the ring.
The radiation stress tensors are given as
\begin{equation}
	(P^{rr},~P^{\varphi\varphi})=\left({E\over 2},~{E \over 2}\right). 
\label{radp}
\end{equation}

\subsection{Radiation Hydrodynamical Equations}

As a relativistic result of radiative absorption 
and subsequent re-emission,
radiation fields exert drag force on moving matter
in resistance to its velocity. 
This radiation drag extracts angular momentum from a rotating disk,
thereby allowing the gas to accrete onto the center.
This is known to be the Poynting-Robertson effect
in the solar system (Poynting 1903; Robertson 1937). 
The radiation drag usually makes effects of $O(v/c)$, but 
it could provide a key mechanism for momentum/angular momentum
transfer in environments of intensive radiation fields
(Loeb, 1993; Umemura, Loeb \& Turner 1993; 
Fukue \& Umemura 1994; Umemura \& Fukue 1994; Tsuribe, 
Fukue, \& Umemura 1994; Tsuribe, Umemura \& Fukue 1995; 
Fukue \& Umemura 1995; Umemura, Fukue \& Mineshige 1997; 
Tsuribe \& Umemura 1997).

Specific radiation forces which are exerted on moving fluid elements 
with velocity \bm{v} are given as
\begin{equation}
f_r={\chi \over c}(F^r-v_r E-v_r P^{rr}-v_\varphi P^{\varphi r} )
\end{equation}
and 
\begin{equation}
f_\varphi={\chi \over c}
(F^\varphi-v_\varphi E-v_\varphi P^{\varphi \varphi}
-v_r P^{r\varphi} ),
\end{equation}
(Mihalas \& Mihalas 1984)
respectively in radial and azimuthal directions.
Here $\chi$ is the mass extinction coefficient, which is
given by $\chi=(\kappa+n_e \sigma_T)/\rho$,
where $\kappa$ is the absorption coefficient, 
$n_e$ is the electron number density,
$\sigma_T$ is the Thomson scattering cross section,
and $\rho$ is the mass density.
Coupled with the radiative quantities presented in the previous section, 
the last term of each component of specific forces turns out to be $O(vV/c^2)$, 
and thus is neglected in the present analyses.
Using equations (\ref{radfp}) and (\ref{radp}),
we have the radiation hydrodynamical 
equations to $O(v/c)$ as
\begin{equation}
{dv_r \over dt}={v_\varphi^2 \over r}-{1\over \rho}{dp \over dr}
               -{d\Phi \over dr}
               +{\chi\over c}\left(F^r-{3\over2}Ev_r \right),\label{dvrdt}
\end{equation}
\begin{equation}
{1 \over r}{d( r v_\varphi) \over dt}={3\chi E\over 2c}
                  \left({r\over R}V-v_\varphi \right),\label{dvpdt}
\end{equation}
where $p$ is the gas pressure, and
$\Phi$ is the external force potential (e.g., gravity).
The azimuthal equation of motion (\ref{dvpdt}) implies that the radiation 
flux force (the first term on the right-hand side) makes fluid elements tend 
to co-rotate with the stellar ring, whereas the radiation drag 
(the second term on the right-hand side) works to extract 
the angular momenta from the elements.
It is noted that both terms are $O(v/c)$
in this equation, in contrast to the radial flux force
of $O(1)$ in equation (\ref{dvrdt}).

\section{Radiative Avalanche}

\subsection{Timescales}

As shown in Paper I, 
the radiation drag time scale is given by
\begin{equation}
	t_\gamma \equiv  {8\pi c^2 R^2 \over 3\chi L_\ast}.
\end{equation}
First, we consider an ionized disk.
The total number per unit time of ionizing photons from starburst 
regions is roughly estimated as 
$\dot{N}_{\rm UV}\sim f_{\rm UV} L_\ast/h\nu_L \sim 5.4 \times 10^{55} {\rm s}^{-1}
(L_\ast / 3 \times 10^{12} L_\odot)$,
with $\nu_L$ being the Lyman limit frequency, and 
$f_{\rm UV}$ is a factor which represents the contribution to UV light
by starbursts (assumed to be $f_{\rm UV}\sim 0.1$ here),
while the total recombination rate to excited levels of hydrogen
is $\dot{N}_{\rm rec} = n_e \alpha^{(2)}M_g/m_p$
$\simeq 1 \times 10^{56} {\rm s}^{-1} (M_g /10^8 M_\odot)
(R_g/{\rm 100 pc})^{-3}$, where $M_g$ and $R_g$ are the mass and the radius 
of the gas disk and
$\alpha^{(2)}$ is the recombination rate coefficient 
to all excited levels of hydrogen.
The optical depth of ionized gas is roughly estimated as
$\tau \sim M_g\sigma_T/\pi R_g^2 m_p$, which is
numerically $\tau \sim 0.26(M_g/10^{8}M_\odot)(R_g/{\rm 100 pc})^{-2}$. 
Thus, photo-ionization could be possible for an optically thin disk or
an optically thin surface layer in an optically thick disk.
For perfectly-ionized gas,
the drag force is equivalent to the Compton drag,
that is, $\chi=\sigma_T/m_p$, where $\sigma_T$ is the Thomson-scattering 
cross-section and $m_p$ is the proton mass. Then, the radiation drag
timescale is estimated as
\begin{equation}
t_\gamma \simeq 4.8\times 10^{9} {\rm yr} 
\left(L_\ast \over 3 \times 10^{12}L_\odot \right)^{-1} 
\left(R \over {\rm 100 pc} \right)^2.
\end{equation}
This is fairly long compared to the duration of AGN phase
or starburst phase inferred from observations ($\sim 10^8$yr).

In practice, a large amount of dust is detected in luminous IR galaxies
(e.g. Scoville et al. 1991).
From theoretical points of view, the timescale of 
photodesorption of dust by UV is longer than the Hubble time
for refractory grains like graphites or silicates, and also
the thermal sputtering rate is completely negligible for such grains
at the temperature of $\sim 10^4$K 
(Burke \& Silk 1974; Draine \& Salpeter 1979). 
If a disk contains dust grains, 
then the accretion timescale could be much shorter
because of its large opacity.
The optical depth is estimated to be
$
\tau \sim 5 \times 10^2
\left( M_g / 10^{8}M_\odot \right)
\left( R_g / {\rm 100 pc}\right)^{-2} 
\left(f_{dg} / 10^{-2}\right),
$
where $f_{dg}$ is the dust-to-gas mass ratio 
(Mathis, Rumpl, \& Nordsiek 1977).
If the relaxation between gas and dust is sufficiently  rapid, then
the dust accretes with accompanying gas.
The relaxation time between gas and dust is estimated by 
\begin{equation}
t_{\rm relax}={3 m_d (2\pi)^{1/2}(kT)^{3/2} \over
8\pi c^4 m_p^{1/2}n_p e^4 Z_d^2 \ln \Lambda}, 
\end{equation}
where $m_d$ is the mass of dust grain, $n_p$ is the number density of
protons, $Z_d$ is the dust charge in units of
proton charge, and $\ln\Lambda$ is the Coulomb logarithm (Spitzer 1962).
Tentatively assuming the thickness of gas disk to be one tenth of
the radius, we evaluate
$
n_p =n_e \simeq 1.3\times 10^{6}{\rm cm}^{-3} 
(M_g / 10^{10}M_\odot) 
(R_g/{\rm 100 pc})^{-3}.
$
Then, for spherical dust grains, 
\begin{equation}
t_{\rm relax}=3.1\times10^{-2}
 {\rm yr} \left({n_p \over 10^6{\rm cm}^{-3}} \right)^{-1}
\left({T \over 10^4{\rm K}} \right)^{-1/2}
\left({a_d \over 0.1 \mu {\rm m} } \right)
\left({\rho_s \over {\rm g~cm}^{-3}} \right)
\left({Z_d \over -40} \right)^{-2},
\end{equation}
where $a_d$ is the grain radius and $\rho_s$ is the density of solid material
within the grain (e.g., Spitzer 1978). 
Hence, dust is tightly coupled with gas in a typical rotation 
timescale of $\sim 10^5$ yr,
and therefore the disk can be regarded to be composed of fluid with 
mass extinction as
\begin{equation}
\chi={n_e\sigma_T + n_d\sigma_d \over \rho_g + \rho_d}, 
\end{equation}
where $n_d$, $\sigma_d$, and $\rho_d$ are the number density, 
cross-section, and mass density of dust, respectively, and $\rho_g$ is 
the mass density of gas.
If taking the dust-to-gas mass ratio of $\sim 10^{-2}$
that is inferred in the Solar neighborhood, then the opacity ratio is
$$
{n_d\sigma_d \over n_e\sigma_T}=1.9\times 10^3 
\left({a_d \over 0.1 \mu {\rm m} } \right)^{-1}
\left({\rho_s \over {\rm g~cm}^{-3}} \right)^{-1}
\left({f_{dg} \over 10^{-2}} \right).
$$
Thus, we find $n_d\sigma_d \gg n_e\sigma_T$ and $f_{dg} \ll 1$
in probable situations. Then, we can evaluate
$\chi \simeq n_d\sigma_d / \rho_g$, so that
\begin{equation}
t_\gamma \simeq 2.4\times 10^{6} {\rm yr}
\left(L_\ast \over 3 \times 10^{12}L_\odot \right)^{-1}
\left(R \over {\rm 100 pc} \right)^2
\left({f_{dg} \over 10^{-2}} \right)^{-1}
\left({a_d \over 0.1 \mu {\rm m} } \right)
\left({\rho_s \over {\rm g~cm}^{-3}} \right). \label{tgamma}
\end{equation}
Hence, for intensive starbursts, the drag timescale
could be shorter than the duration of AGN phase or starburst phase.

\subsection{Disk Evolution}

Suppose that an initially quiescent gas disk is abruptly ignited
by a starburst at the radius $R$ with luminosity $L_\ast$.
Since the radiation drag efficiency decreases exponentially 
with the optical depth (Tsuribe \& Umemura 1997),
just an optically thin surface layer avalanches toward a nucleus.
In equation (\ref{dvrdt}), the radial radiation flux force, 
$\chi F^r/c$, is obviously greater by $O(c/v)$ than the drag force.
So, the radiation drag makes a negligible contribution in
the {\it radial} equation of motion. On the other hand, 
in azimuthal directions (see equation [\ref{dvpdt}]), 
both the radiation flux force and the radiation drag are $O(v/c)$,
and therefore they work simultaneously.
Thus, the disk evolution is expected to proceed in two steps:
First, the optically thin surface layer 
(or the whole of an optically thin disk) 
is contracted by the radial radiation flux force and
is quickly settled in quasi-rotation balance. Then, 
the layer sheds angular momentum due to the radiation drag,
and consequently undergoes an avalanche.

With the radial equation of motion (\ref{dvrdt}),
disregarding pressure gradient force as well as the drag force, 
the condition of rotation balance is given by
\begin{equation}
{v_\varphi^2 \over r}={d\Phi \over dr}-{\chi\over c}F^r.\label{rotb}
\end{equation}
Here, we assume a gravitational potential such that
\begin{equation}
r{d\Phi \over dr} \equiv V^2 \left( {r \over R}\right)^{2n}.
\label{rotl}
\end{equation}
The rotation velocity of the stellar ring 
is given by $V=(GM_{\rm dyn}/R)^{1/2}$, where $M_{\rm dyn}$ is the 
dynamical mass contained within $R$.
If $F^r$ is negligibly small, 
then $n=-1/2$ simply corresponds to Keplerian rotation, 
$n=0$ to flat rotation, and $n=1$ to rigid rotation. 
Equation (\ref{rotb}) is
rewritten with (\ref{rotl}) into a dimensionless form as
\begin{equation}
\tilde{v}_\varphi^2
=x^{2n}+\Gamma x^2\left[ {1 \over 2} + {x^2 \over \pi (1-x)}\right],
\label{vp2}
\end{equation}
with using
$x\equiv r/R$, $\tilde{v}_\varphi\equiv v_\varphi/V$, and
$\Gamma \equiv L_\ast/L_{\rm E}$, where $L_{\rm E}$ is
the Eddington luminosity corresponding to $M_{\rm dyn}$,
\begin{equation}
L_{\rm E}={4\pi cGM_{\rm dyn} \over \chi}.
\end{equation}
For a dusty disk, the Eddington luminosity is numerically given as
\begin{equation}
L_{\rm E}=1.68 \times 10^{11} L_\odot 
\left( {M_{\rm dyn} \over 10^{10} M_\odot}\right) 
\left({f_{dg} \over 10^{-2}} \right)^{-1}
\left({a_d \over 0.1 \mu {\rm m} } \right)
\left({\rho_s \over {\rm g~cm}^{-3}} \right). 
\end{equation}
The azimuthal equation of 
motion (\ref{dvpdt}) can be also rewritten in a dimensionless form as
\begin{equation}
{d(x \tilde{v}_\varphi)\over d\tilde{t}}
={x(x-\tilde{v}_\varphi)\over (1-x^2)},
\label{djdt}
\end{equation}
where $\tilde{t}\equiv t/t_\gamma$.

To begin with, we see the first stage, i.e., the radial contraction phase. 
As shown in equation (\ref{vp2}), 
the radial compression by the flux force is expected to be strongest
on the outer edge of disk.
Because the radial flux does not extract angular momentum,
the angular momentum conservation determines the radius of rotation
equilibrium. Since $\tilde{v}_\varphi=1$
at the outer edge of the disk before a starburst, 
the equilibrium radius after a starburst is determined 
by the condition of $\tilde{v}_\varphi=1/x$.
Combined with equation (\ref{vp2}),
the resultant equilibrium radii $r_{\rm eq}$ for the material that is 
originally located at the outer edge
are shown versus $\Gamma$ for three types of
rotation potentials in Figure 1. The $r_{\rm eq}$ is
a slowly decreasing function with $\Gamma$ because of
angular momentum barrier. 
The asymptotic behavior for $\Gamma \ll 1$ is
$r_{\rm eq}/R=1-[\Gamma/2\pi(n+1)]^{1/2}$, while
$r_{\rm eq}/R=(2/\Gamma)^{1/4}$ for $\Gamma \gg 1$.
Equation (\ref{vp2}) shows that the radial radiation flux enhances the
rotation velocity, so that a shear flow may be excited on the boundary
between the optically thin surface layer and 
the optically thick deep stratum.
The fractional difference of rotation velocity,
which is defined by
$\Delta v_\varphi /v_\varphi \equiv (\tilde{v}_\varphi-x^n)/x^n$,
is also shown at $r_{\rm eq}$ in Figure 1.
$\Delta v_\varphi /v_\varphi$ is found to be 
weakly dependent upon $\Gamma$, and 
we have asymptotic solutions as 
\begin{equation}
{\Delta v_\varphi \over v_\varphi}=\left\{
\begin{array}{ll}
{\displaystyle \left[{(n+1) \Gamma \over 2\pi} \right ]^{1/2}}
& \cdots \Gamma \ll 1 \\
{\displaystyle \left({\Gamma \over 2} \right)^{n+1\over 4}}
& \cdots \Gamma \gg 1.
\end{array}
\right.
\end{equation}
If $\Gamma =O(1)$ (which is indeed the case for active galaxies), 
$\Delta v_\varphi /v_\varphi=0.2-0.6$ in the range of
$n=-1/2$ to 1. This shear flow may generate turbulence 
due to the Kelvin-Helmholtz instabilities, or the viscosity could
thermalize the shear flow energy.

Next, we consider the avalanche phase.
If the radiation is very weak ($\Gamma \ll 1$), 
equation (\ref{djdt}) coupled with (\ref{vp2}) 
is reduced to a differential equation as
\begin{equation}
x'\equiv {dx \over d\tilde{t}} 
={(1-x^{n-1}) \over (n+1)x^{n-2}(1-x^2)}. \label{dxdt0}
\end{equation}
This equation can be analytically solved for some values of $n$.
For instance, the Keplerian rotation potential ($n=-1/2$) leads to the solution 
\begin{equation}
\tilde{t}=-\sqrt{x}+{1 \over \sqrt{3}} 
\tan^{-1}\left(1+2\sqrt{x} \over \sqrt{3}\right)
-{1 \over 2}\log \left(x \over 1+x+\sqrt{x} \right)-0.154.
\end{equation}
If we take a Mestel disk ($n=0$), then the solution is
\begin{equation}
\tilde{t}=1-x-\log x.
\end{equation}
The rigid-rotation potential results in $x$=const. In other words, 
no accretion is driven in a rigid-rotation potential 
under the circumstances of very weak radial radiation flux 
(see also Paper I).
If $\Gamma=O(1)$, however,
the effects of radial flux force is important not only in radial 
contraction but also in the angular momentum transfer.
When the radial flux force is included, the accretion is governed by
\begin{equation}
x'={\tilde{v}_\varphi(x-\tilde{v}_\varphi) 
\over (1-x^2) h(x)},
 \label{dxdt}
\end{equation}
from equations (\ref{vp2}) and (\ref{djdt}),
where
\begin{equation}
h(x)= (n+1)x^{2n-1}+\Gamma x \left[1+{ x^2(6-5x) \over 2\pi(1-x)^2} \right].
\end{equation}
As shown below soon, this equation allows the mass accretion even
in a rigid-rotation potential.

\subsection{Mass Accretion Rate}

If we suppose a geometrically thin, optically thick disk, then 
the upper surface is irradiated by radiation from 
the upper half of starburst regions and the lower surface is
irradiated by the lower half.
Taking this into account, we can estimate the mass accretion rate as
\begin{eqnarray}
\dot{M} & \equiv & - \int 2\pi r v_r \rho dz \nonumber \\
 & \simeq & -(1+{\rm e}^{-\tau/2})\int_0^{\tau/2} 
{2\pi r v_r \over \chi} {\rm e}^{-\tau_1}d\tau_1,
\end{eqnarray}
where $z$ is the vertical coordinate, $d\tau_1 \equiv \chi \rho dz$,
and $\tau$ is the total face-on optical depth.
This can be rewritten in a dimensionless form using the same notations 
as the previous subsection,
\begin{equation}
\dot{m}_\gamma=-{3 \over 4}(1+{\rm e}^{-\tau/2})\int_0^{\tau/2} 
xx'{\rm e}^{-\tau_1}d\tau_1
=-{3 \over 4}xx'(1-{\rm e}^{-\tau}),
\end{equation}
where $\dot{m}_\gamma \equiv \dot{M} /(L_\ast/c^2)$ and $x'$ 
is given by (\ref{dxdt}).
In an optically thick limit, 
\begin{equation}
\dot{m}_\gamma=-{3 \over 4}xx'. \label{mdot}
\end{equation} 
Hence, the mass accretion rate is independent of the extinction
coefficient as well as the density distribution in an optically 
thick disk. In comparison, in an optically thin regime,
$
\dot{m}_\gamma=-3xx'\tau/4.
$
This implies that the mass accretion rate is smaller in an optically thin disk
than in an optically thick disk. This is because 
in optically thin media a number of photons penetrate the gas disk
without collision with matter, and hence not all the photons from 
the starburst regions are available for radiation drag.

The fact that the radial flux raises the rotational velocity
brings two effects on the mass accretion rate 
which mutually compete: One is a positive effect that
the radiation drag force becomes more efficient due to enhanced $v_\varphi$,
and the other is negative effect that the enhanced $v_\varphi$ leads to
the increase of specific angular momentum at fixed $r$, 
so that the accretion is decelerated.
In Figure 2, for a wide variety of $\Gamma=L_\ast/L_{\rm E}$,
the resultant mass accretion rate is shown as a function of
radii in an optically thick regime.
In this figure, a thin solid curve shows the estimation
in the case of the negligible effects of the radial flux, i.e., 
\begin{equation}
\dot{m}_\gamma={3 \over 8} { 1-n \over 1+n } x^2 \label{mdot00}
\end{equation}
(Paper I). The thick curves are the present results, where filled circles
represent the equilibrium radii of outer-edge shown in Figure 1.
As shown in Figure 2a, for a Keplerian disk,
the negative effect overwhelms the positive one 
around the outer-edge of the disk.
On the other hand, the positive effect is predominant for a Mestel-like disk 
($n=0$) as shown in Figure 2b. 
In the vicinity of the ring ($x \rightarrow 1$),
we have an asymptotic solution as $\dot{m}_\gamma=3/4$, 
being independent of $\Gamma$ as well as $n$. 
In the opposite limit, i.e., $x \rightarrow 0$, the solutions are
not dependent on $\Gamma$ for $-1<n<1$,
and then the mass accretion rate is basically determined
by (\ref{mdot00}).

It is worth stressing that 
the radial flux can allow the accretion
even if the disk is governed by a rigid-rotation potential ($n=1$).
As shown in Figure 2c, the induced mass accretion is not sensitive to
$\Gamma$, and the maximal rate is $\dot{M}\sim 0.2 L_\ast/c^2$
in the range of $\Gamma=10^{-3}$ to $10^2$.
At small radii, there is an asymptotic solution as
\begin{equation}
\dot{m}_\gamma={3 \over 8}x^2 \left[1-\left(1+{\Gamma \over 2} 
\right)^{-1/2} \right].
\end{equation}
This approaches a limiting value $3x^2/8$ for 
$\Gamma \rightarrow \infty$, which coincides with (\ref{mdot00})
for $n=0$.

In the long run, the mass accretion rate has turned out to be
almost insensitive to the rotation law and given by
\begin{eqnarray}
\dot{M} &= &\eta \left(L_\ast \over c^2 \right) x^2
(1-{\rm e}^{-\tau}) \nonumber \\ 
 & \sim & 0.2M_\odot {\rm yr}^{-1} ~\eta 
\left( L_\ast \over 3 \times 10^{12} L_\odot \right)
x^2 (1-{\rm e}^{-\tau}),
\label{mdot0}
\end{eqnarray}
where $\eta \simeq 0.2-1$.
Such radial dependence as $x^2$ implies that
an avalanche occurs in a homologous manner.
The radial dependence slightly deviates from $x^2$ near the outer edge
of the disk. Especially, it is faster than $x^2$ in a rigid-rotation
potential, so that some amount of mass would accumulate
around $r_{\rm eq}$.

Although we have assumed a ring-like configuration for starburst regions,
the regions may be extended over a wide range of radii and heights.
However, equation (\ref{rade}) shows that
the radiation from the inner edge of the star-forming regions would
be most contributive
as long as the surface brightness falls faster than
$\Sigma_L(R) \propto (1-r^2/R^2)$, which is basically constant for $R \gg r$.
In other words, the regions of peak surface brightness would be always
most responsible for the radiative avalanche.
Also, in an optically thick regime,
the effects of viewing-angles from starburst regions to a disk
might reduce the mass accretion rate, depending upon
the configuration of starburst regions (Taniguchi 1997).
If we suppose a torus of emission regions with the half thickness 
of $\Delta R$,
then the rate should be multiplied roughly by $\Delta R/R$
due to a geometrical dilution effect (Ohsuga et al. 1997).

\section{Stability of Disk}

\subsection{Radiation-Hydrodynamical Instability}

If we suppose a geometrically thin, optically thick disk,  
the upper surface is irradiated downward, while
the lower surface is irradiated upward.
Hence, the disk is rather confined by radiation force. 
But, the outer edge of an optically
thick disk (of course, an optically thin disk as well) 
could be vertically unstable, especially if the starburst luminosity
is super-Eddington.
As far as a Kepler disk or a Mestel disk is concerned, 
the gravity is always dominant at small radii [see (\ref{radfr})].
Thus, the inner disk remains stable,
whereas the outer edge could be blown away by radiative acceleration.
The radius inside which the disk is stable
is given by the root of an equation as
\begin{equation}
x^{2n-3}(1-x)=\Gamma\left[ {1 \over 2} + {x^2 \over \pi(1-x) }\right].
\end{equation}
The solutions are shown in Figure 3 for $n=-1/2$, 0, and 1.
We find that the radius is sensitively dependent upon $\Gamma$
if $\Gamma \gsim 1$. 
To see the fate of the blown-out gas in three-dimensional space,
we compare the radiation flux force with the gravitational force
in Figure 4, illustratively for $n=-1/2$.
On the hatched sides for each value of $\Gamma$, 
the radiative force overwhelms the gravity in the vertical directions.
The dotted lines are corresponding to the unstable branches below which
gas is accelerated downward by the gravity, while above which
gas blows upward due to radiative force.
The solid lines are stable branches. 
When $\Gamma>2$, 
no stable branch exists and therefore the blown-out gas may result in
a superwind (a ``blizzard'' phase).
[If we assume the initial mass function which is 
deficient in low-mass stars as observed in starburst regions (see also \S 4.1),
$\Gamma>2$ is corresponding to early stages of $<5\times 10^7$yr 
after the coeval star-formation. ]
If the gravity by the bulge stars significantly decelerates the wind, 
the spurted-out gas may fall onto the inner disk 
with losing angular momentum by a fraction 
of $\sim R/Vt_\gamma$ (cf. Meyer \& Meyer-Hofmeister 1994). 
When $\Gamma<2$, the stable branch
composes a wall surrounding the starburst ring.
In particular, the wall becomes of torus shape if $\Gamma<1$.
Taking account of the fact that the radiation force becomes 
ineffective for optically-thick gas, 
the optical depth of the formed wall should be of $O(1)$.
The radiation emitted originally in UV is absorbed by this wall and re-emitted
possibly in IR wavelength. However, the radiative avalanche is almost independent
of the spectrum of the radiation, but just depends upon the
bolometric luminosity. Hence, the emission from the wall could 
induce an indirect radiative avalanche.
Also, such a radiatively built-up obscuring 
wall may diminish the geometrical dilution, although
the direct avalanche may be strongly subject to the geometrical dilution
if the thickness of the starburst regions is as small as 10pc
(Taniguchi 1997).

The radiatively built-up obscuring wall may 
be a different version of the formation of an obscuring torus
from those proposed so far 
(e.g. Krolik \& Begelman 1988; Yi, Field, \& Blackman 1994).
In the present mechanism, the angular extension of the obscuring wall
is a decreasing function of time according as the starburst luminosity 
fades out. Thus, the classification of AGN types by anisotropic viewing
could be comprehended
in the contexts of the evolution of circumnuclear starbursts. 
The further detailed confrontation with observations will be made elsewhere.

Finally, even an inner gravity-dominant disk could be unstable to {\it warping} 
through a similar mechanism proposed by Pringle (1996, see also
Maloney et al. 1996), although
the radiative acceleration is reduced by a factor of $\sim 1/\tau$.
If a radiative backreaction torque can drive a warp, the illumination
to the warped disk is augmented, thereby increasing the accretion.

\subsection{Gravitational Instability}

We consider three components of gravity sources, that is, a central
massive black hole, a massive disk surrounding
the black hole, and the galactic bulge. 
The surface density distribution of the disk is assumed to be 
Mestel-like, and the bulge is to be uniform.
To see the stability, we evaluate the Toomre's $Q$ to find
\begin{equation}
Q={2c_s \over V f_g}\left( {f_{\rm BH}\over x}
+2f_g +4f_{\rm bulge} x^2 \right)^{1/2},
\end{equation}
where $V=(GM_{\rm dyn}/R)^{1/2}$ is the virial velocity for
the total dynamical mass $M_{\rm dyn}$ within $R$, and
$f_{\rm BH}$, $f_g$ and $f_{\rm bulge}$ are respectively
the mass fraction of the black hole, the disk, and
the bulge to the whole system. 
Here, we have assumed that the disk radius is 
comparable to $R$. 
If the gravity of the black hole or the bulge is predominant,
then $Q>1$ as readily seen, so that the disk can be stabilized.
However, if the cooling is effective to give $c_s \ll V$,
then the disk can be gravitationally unstable to a high degree.
In some cases, the presence of an appreciable velocity dispersion
could help to weaken the instability of the disk.
For instance, the nuclear disk in Arp 220 is stabilized in outer
regions of $\gsim 400$pc due to a large velocity dispersion of 
$\sim 90$km s$^{-1}$, while the inner region of $\lsim 400$pc
is marginally unstable (Scoville, Yun, \& Bryant 1997).
Anyhow, if the disk is unstable in some regions, it can fragment 
into lumps, presumably forming stars.
Since the radiation drag works just as a surface effect for
optically thick lumps, the effects of the drag force would be negligible
for them. Nonetheless, if at least a portion of $1/\tau$ is left 
in the form of diffuse gas (it is not unlikely because $\tau \gsim 10^3$),
it may be subject to effective drag.
Even if all the matter 
was processed into stars, the stellar mass loss would supply
diffuse gas. Also, the newly formed stars may exert
radiative force on the innermost gas disk.

\section{Discussion}

\subsection{Sequential Starbursts}

Starbursts might not be coeval, but could be sequential inward. 
There are a lot of possibilities for the sequential starbursts.
The radial radiation flux can compress the outer edge
of gas disk, where star formation may be triggered. 
If the UV radiation produces the D-type ionization front at the outer edge,
massive stars could form there (Elmegreen \& Lada 1977).
If sequential explosions of supernovae 
occur in an OB association embedded in a plane-stratified gas disk,
they build up an elongated hot cavity ({\it superbubble}), and consequently
the shock-heated gas spurts out from the disk 
(Shapiro \& Field 1976; Tomisaka \& Ikeuchi 1986; 
Norman \& Ikeuchi 1989). 
The superbubble may also compress the outer edge of the disk by
shocks, and new stars may be born there.
Furthermore,
the successive interaction of starburst winds with the inner gas disk
might be possible through a similar mechanism to that 
considered by Umemura \& Ikeuchi (1987).
Anyhow, if an compressed gas ring is ignited by star-formation, 
it may induce a further radiative avalanche on the inner gas disk.

\subsection{Effects of Stellar Evolution}

Recently, it is reported that in starburst regions
the stellar initial mass function (IMF) is deficient in low-mass stars,
with the cutoff of about $m_{\rm l}=2M_\odot$, and the upper mass 
limit is inferred to be around $m_{\rm u}=40M_\odot$
(Doyon, Puxley, \& Joseph 1992; Charlot et al. 1993; 
Doane \& Mathews 1993; Hill et al. 1994; Blandle et al. 1996).
Here, we consider the effects of stellar evolution. 
We assume a Salpeter-type IMF for a mass range of 
$[m_{\rm l},~m_{\rm u}]$ as
$
\phi (m_\ast)=\phi_0 (m_\ast / M_\odot)^{-(1+s)},
$
where $s=1.35$ and $m_\ast$ is mass of a star.
We employ simple relations of the stellar 
mass to luminosity as
$
(l_\ast/L_\odot)=(m_\ast / M_\odot)^q 
$ 
with $q=3.7$, 
and the mass to age as
$
t_\ast=1.1\times 10^{10}{\rm yr}(m_\ast / M_\odot)^{-\omega}
$
with $\omega=2.7$ (Lang 1974). Then, we obtain 
the total stellar mass and luminosity 
respectively as $M_\ast=1.46\phi_0 M_\odot$ and
$L_\ast =2.2\phi_0 (82t_7^{-0.87}-1)L_\odot$, 
where $t_7$ is the elapsed time in units of $10^7$yr after  
the ignition of a starburst.  These give a large
luminosity-to-mass ratio in early stages of evolution as
$L_\ast/M_\ast \simeq 130 t_7^{-0.87}(L_\odot/M_\odot)$, 
Also, if we assume that stars of $>4M_\odot$
are destined to undergo supernova explosions and
release the energy radiatively with the efficiency of $\epsilon_{\rm SN}$
to the rest mass energy, the temporally averaged supernova luminosity is 
$L_{\rm SN}=\epsilon_{\rm SN}\dot{M}_{\rm SN}c^2
=20\phi_0t_7^{-0.87} L_\odot$, where $M_{\rm SN}(t)$ is the total mass
of SNs which exploded until $t$, and $\epsilon_{\rm SN}=10^{-4}$ is
assumed. Thus, the ratio of $L_{\rm SN}$ to $L_\ast$ is 
$L_{\rm SN}/L_\ast=0.12/(1-0.01t_7^{0.87})$. Hence, 
the luminosity is always dominated by stellar luminosity until 
the supernova explosions are terminated at $t_7=26$.

\subsection{Generation of Seed Magnetic Fields}

Although the Compton drag is not predominant in a dusty disk, 
it is significant in the sense that magnetic fields could be generated,
because the Compton drag force is exerted virtually on free electrons
(Mishustin \& Ruzmaikin 1971; Zeldovich, Ruzmaikin \& Sokoloff 1983).
The equation of motion for electrons is
\begin{equation}
-(e/m_e c)(c\bm{E}+\bm{v}_e\times\bm{B})-
\nu_{e\gamma}\bm{v}_e+\nu_{ep}(\bm{v}_p-\bm{v_e})=0,
\end{equation}
where $m_e$ is the electron mass,
$\bm{v}_e$ and $\bm{v}_p$ are the electron and proton velocities,
$\bm{E}$ and $\bm{B}$ are the induced electric and magnetic fields,
$\nu_{e\gamma}= m_p/m_e t_\gamma=3\sigma_T L_\ast/8\pi m_e c^2 R^2$,
and $\nu_{ep}=4.8 n_e T^{-3/2} \ln\Lambda$ (s$^{-1}$)
with $\ln\Lambda$ being
the Coulomb logarithm ($\simeq 30$). 
Since $\nu_{e\gamma}$ is estimated as
$
\nu_{e\gamma} \simeq 1.2\times 10^{-14}{\rm s}^{-1} 
(L_\ast / 3 \times 10^{12}L_\odot) 
(R_g/{\rm 100 pc})^{-2},
$
$\nu_{e\gamma} \ll \nu_{ep}$, and therefore
$\bm{v}_e$ is approximately equal to the fluid velocity 
($\bm{v}_e \simeq \bm{v}$).
Then, Maxwell equations give
\begin{equation}
{\partial \bm{B} \over \partial t}=
{m_e c \nu_{e\gamma} \over e}\nabla \times \bm{v}
-{m_e c^2 \nu_{ep} \over 4\pi n_e e^2}\nabla^2\bm{B}.
\end{equation}
Here, one may ignore the second term of the right-hand side for fields of
galactic scale, because it arises from the Coulomb interactions
(Zeldovich, Ruzmaikin \& Sokoloff 1983). Thus,
$\bm{B}$ is evaluated by
\begin{equation}
\bm{B}={ m_e c \over e} \int \nu_{e\gamma} \nabla \times \bm{v} dt.
\end{equation}
If the gas disk is in Kepler rotation, then 
$|\nabla \times \bm{v}| \sim v_\varphi /r$, so that
\begin{equation}
|\bm{B}| \simeq 4.6\times 10^{-13}{\rm Gauss} 
\left({L_\ast \over 3 \times 10^{12}L_\odot} \right) 
\left({R \over {\rm 100 pc}} \right)^{-2}
\left({t_{SB} \over 10^8 {\rm yr}} \right)
\left({M_{\rm dyn} \over 10^{10}M_\odot} \right)^{1/2}
\left({r \over {\rm pc}} \right)^{-3/2}.
\end{equation}
Remembering that the radial radiation flux 
produces shear on the disk surface,
these seed magnetic fields could be quickly enhanced due to 
the combination of differential rotation, turbulence, and
magneto-rotational shear instabilities (Balbus \& Hawley 1991;
Hawley, Gammie \& Balbus 1995; Matsumoto \& Tajima 1995; 
Matsumoto et al. 1996), which may be relevant to inner viscous flow.

\subsection{Link to an Advection-Dominated Flow}

Recently, the solution of an advection-dominated viscous flow
has been developed in the prescription of $\alpha$-viscosity 
(Abramowicz et al. 1995;
Chen et al. 1995; Narayan \& Yi 1994, 1995; Narayan 1996;
Narayan, Kato, \& Honma 1997; Chen, Abramowicz, \& Lasota 1997). 
In the present radiation-hydrodynamical model,
the strong shear flow which is comparable to rotation velocity 
is induced on the disk surface, and hence a hot corona
with roughly virial temperature may be generated. This is a favorable 
situation where an advection-dominated accretion is raised.
The advection-dominated solution exists when the mass accretion rate
is relatively low as $\dot{M} \lsim \dot{M}_{\rm crit}$.
If a fiducial efficiency of 10\% is assumed
(Frank, King, \& Raine 1992), then the critical rate is given as
$\dot{M}_{\rm crit} \simeq 3 \alpha^2 L_{\rm E,Th}
(r/100r_g)^{-1/2}/c^2$ at $r \geq 100r_g$ and constant at $r < 100r_g$, 
where $\alpha$ is the viscosity parameter 
($\alpha <1$, Shakura \& Sunyaev 1973), $r_g$ is the Schwarzshild radius
of the central black hole and
$L_{\rm E,Th}$ is the Eddington luminosity corresponding to
the Thomson scattering. 
The energy conversion efficiency is 
$\epsilon \equiv L_{\rm adv}/\dot{M}c^2 \sim 0.1(\dot{M}/\dot{M}_{\rm crit})$,
where $L_{\rm adv}$ is the energy release 
from an advection-dominated accretion 
onto a putative massive black hole. When equating $\dot{M}_{\rm crit}$ with
(\ref{mdot0}), we find a transition radius where the mass accretion
by the radiative avalanche links to an advection-dominated flow. 
The transition radius is given by
\begin{eqnarray}
x_{\rm tr} \equiv {r_{\rm tr} \over R} 
 &=& 2.5 \times 10^{-2} \eta^{-2/5} 
\left(\alpha \over 0.1 \right)^{4/5}
\left( M_{\rm BH} \over 10^{8}M_\odot \right)^{3/5}
\left( L_\ast \over 3\times 10^{12}L_\odot \right)^{-2/5}
\left( R \over 100 {\rm pc}\right)^{-1/5},
\end{eqnarray}
where $M_{\rm BH}$ is the mass of a central black hole.
[It is noted that more precisely
we should employ the total dynamical mass in relevant scales 
instead of $M_{\rm BH}$ (Mineshige \& Umemura 1996, 1997),
although $M_{\rm dyn}\sim M_{\rm BH}$ if $x_{\rm tr}\ll 1$.]
Thus, if $R\sim100$pc, the radiative avalanche may link to 
an advection-dominated accretion around $\sim1$pc.

\section{Conclusions}

We have investigated fundamental physical processes of
radiatively driven mass accretion due to
circumnuclear starbursts by solving the radiation-hydrodynamical
equations of motion 
including radiation drag force as well as
radiation flux force. The present results are summarized as follows.

\begin{enumerate}

\item 
The mass accretion rate via a radiative avalanche
is almost insensitive to a gravitational potential, 
if the starburst luminosity is 
comparable to or greater than the Eddington luminosity.
In particular, the radial flux force allows the disk gas to accrete
even in a rigid-rotation potential.
The mass-accretion rate at radius $r$ is given by
$\dot{M}(r)= \eta (L_\ast / c^2)(r/R)^2 (\Delta R/R)(1-{\rm e}^{-\tau}) $,
where $\eta$ ranges from 0.2 up to 1, $L_\ast$ and $R$ are  respectively
the bolometric luminosity and the radius of a starburst ring, 
$\Delta R$ is the extension of the emission regions, and $\tau$ is
the face-on optical depth of a disk.
In an optically thick regime, 
the rate depends upon neither the optical depth nor
surface mass density distribution of a disk.
The timescale of the radiative avalanche could be shorter than
$10^8$yr for a dusty disk. 

\item
The radiation-hydrodynamical instability could prevent
the disk surface from accretion especially in early stages 
($\lsim 5 \times 10^7$ yr) of stellar evolution. 
In later stages, it would result in forming
an obscuring wall of torus shape. The wall indirectly induces
a radiative avalanche with diminishing the geometrical dilution 
($\Delta R/R$).
Also, such a radiatively built-up wall shades the nucleus in edge-on views, 
so this is another version of the formation of an obscuring torus.
In the present mechanism, the angular extension of the obscuring wall
is a decreasing function of time according as the starburst luminosity 
fades out. Thus, the classification of AGN types by anisotropic viewing
could be comprehended in the contexts of the evolution of circumnuclear 
starbursts. 
This picture will be confronted with observations in further detail 
elsewhere.

\item 
For an ionized surface of disk, since
the Compton drag works virtually on free electrons,
magnetic fields could be generated. 
Although the magnitude is as small as
an order of $\sim 10^{-12}$ Gauss,
it could be amplified quickly by 
the combination of differential rotation, turbulence, and
magneto-rotational instabilities, which
may be relevant to the inner viscous flow. 

\item
Since the strong shear flow induced by radiation force would result
in hot coronae with roughly the virial temperature,
a radiative avalanche is likely to link to an advection-dominated accretion
at a point where the mass accretion rate falls to
the critical rate, e.g., $\lsim 1$pc.
The link to an advection-dominated flow is possibly related to
X-ray properties in ultra-luminous {\it IRAS} galaxies.

\end{enumerate}

\noindent
{\bf ACKNOWLEDGEMENTS}

\vspace{2mm}
We are grateful to Drs. N. Arimoto, 
T. Nakamoto, R. Narayan, Y. Taniguchi, E. L. Turner,  
T. Yamada, and I. Yi for stimulating discussion.
We also thank the anonymous referee for valuable comments. 
This work was carried out at the Center for Computational Physics 
of University of Tsukuba 
with support from the Japan-US Cooperative Research
Program which is funded by the Japan Society for the Promotion of 
Science and the US National Science Foundation.
Also, this work was supported in part by the Grants-in Aid of the
Ministry of Education, Science, and Culture, 06640346.

\newpage

\noindent
{\large\bf REFERENCES}

\vspace{2mm}
\re
Abramowicz, M. A., Chen, X., Kato, S., Lasota, J-P, \& Regev, O. 
1995, ApJ, 438, L37
\re
Bahcall, J. N., Kirhakos, S., Saxe, D. H., \& Schneider, D. P. 1997, 
ApJ, 479, 642
\re
Balbus, S. A., \& Hawley, J. F. 1991, ApJ, 376, 214
\re
Barth, A. J., Ho, L. C., Filippenko, A. V., \& Sargent, W. L. W.1995, AJ, 110, 1009
\re
Burke, J. R., \& Silk, J. 1974, ApJ, 190, 1
\re
Buta, R., Purcell, G. B., \& Crocker, D. A. 1995, AJ, 110, 1588
\re
Charlot, S., Ferrari, F., Mathews, G. J., \& Silk, J. 1993, ApJ, 419, L57
\re
Chen, X., Abramowicz, M. A., \& Lasota 1997, ApJ, 476, 61
\re
Chen, X., Abramowicz, M. A., Lasota, J-P, Narayan, R., \& Yi, I. 
1995, ApJ, 443, L61
\re
Doane, J. S., \& Mathews, W. G. 1993, ApJ, 419, 573
\re
Doyon, R., Puxley, P. J., \& Joseph, R. D. 1992, ApJ, 397, 117
\re
Draine, B. T., \& Salpeter, E. E. 1979, ApJ, 2331, 438
\re
Elmegreen, B. G., \& Lada, C. J. 1977, ApJ, 214, 725
\re
Ferrarese, L., Ford, H. C., \& Jaffe, W. 1996, ApJ, 470, 444
\re
Fillipenko, A. V.
1992, in Relationships Between Active Galactic Nuclei and
Starburst Galaxies, ed. Fillipenko, A. V.
(ASP Conf. Ser. 31, San Francisco, 1992), 253
\re
Forbes, D. A., Norris, R. P., Williger, G, M., \& Smith, R. C. 
1994, AJ, 107, 984
\re
Frank, J., King, A. \& Raine, D. 1992, Accretion Power in Astrophysics
(Cambridge: Cambridge Univ. Press)
\re
Fukue, J., \& Umemura, M. 1994, PASJ, 46, 87
\re
Fukue, J., \& Umemura, M. 1995, PASJ, 47, 429
\re
Hawley, J. F., Gammie, C. F., \& Balbus, S. A. 1995, ApJ, 440, 742
\re
Heckman, T. M. 1991, 
in Massive Stars in Starburst Galaxies, 
ed. Leitherer, C., Walborn, N., Heckman, T., \& Norman, C.
(Cambridge Univ. Press, Cambridge), 289
\re
Hill, J. K., Isensee, J. E., Cornett, R. H., Bohlin, R. C., O'Connell, R. W.
Roberts, M. S., Smith, A. M., \& Stecher, T. P. 1994, ApJ, 425, 122
\re
Hooper, E. J., Impey, C. D., \& Foltz, C. B. 1997, ApJ, 480, L95
\re
Jaffe, W., Ford, H. C., Ferrarese, L. A., Vandenbosch, F., \& O'Connell, R. W.
1993, Nature, 364, 213
\re
Krolik, J. H., \& Begelman, M. C. 1988, ApJ, 329, 702
\re
Lang, K. R. 1974, in Astrophysical Data: Planets and Stars 
(Springer-Verlag)
\re
Leitherer, C., Vacca, W. D., Conti, P. S., Filippenko, A. V., Robert, C, \&
Sargent, W. L. W.  1996, ApJ, 465, 717
\re
Loeb, A. 1993, ApJ, 403, 542
\re
Maloney, P. R., Begelman, M. C., and Pringle, J. E. 1996, ApJ, 472, 582
\re
Maoz, D., Barth, A. J., Sternberg, A., Filippenko, A. V., Ho, L. C. 
Macchetto, F. D., Rix, H.-W., \& Schneider, D. P. 1996, AJ, 111, 2248
\re
Mathis, J. S., Rumpl, W., \& Nordsiek, K. H. 1977, ApJ, 217, 425
\re
Matsumoto, R., \& Tajima, T.  1995, ApJ, 445, 767
\re
Matsumoto, R., Uchida, Y., Hirose, S., Shibata, K., Hayashi, M. R.,
Ferrari, A., Bodo, G., \& Norman, C. 1996, ApJ, 461, 115
\re
Mauder, W., Weigelt, G., Appenzeller, I., \& Wagner, S. J. 1994, A\&A, 285, 44
\re
McLeod, K. K., \& Rieke, G. H. 1995, ApJ, 454, L77
\re
Mestel, L. 1963, MNRAS, 126, 553
\re
Meyer, F., \& Meyer-Hofmeister, E. 1994, A\&A, 288, 175
\re
Mihalas, D., \& Mihalas, B. W. 1984, in Foundations of
Radiation Hydrodynamics (Oxford Univ. Press) 
\re
Mineshige, S., \& Umemura, M. 1996, ApJ, 469, L49
\re
Mineshige, S., \& Umemura, M. 1997, ApJ, 480, 167
\re
Mishustin, I. N., \& Ruzmaikin, A. A. 1971, Zh. Eksp. Teoret. Fiz, 61, 441
\re
Narayan, R. 1996, ApJ, 462, 136
\re
Narayan, R., Kato, S. \& Honma, F. 1997, ApJ, 476, 49
\re
Narayan, R., \& Yi, I. 1994, ApJ, 428, L13
\re
Narayan, R., \& Yi, I. 1995, ApJ, 452, 710
\re
Norman, C. A., \& Ikeuchi, S. 1989, ApJ, 345, 372
\re
Norman, C. A., \& Scoville, N. 1988, ApJ, 332, 124
\re
Ohsuga, K., Umemura, M, Fukue, J., \& Mineshige, S. 1997, ApJ, submitted
\re
Ohta, K., Yamada, T., Nakanishi, K., Kohno, K.,
 Akiyama, M., \& Kawabe, R. 1996, Nature, 382, 426
\re
Omont, A., Petitjean, P., Guilloteau, S., McMahon, R.G.,
 Solomon, P.M., \& P\'{e}contal, E. 1996, Nature, 382, 428
\re
Perry, J. J. 
1992, in Relationships Between Active Galactic Nuclei and
Starburst Galaxies, ed. Fillipenko, A. V.
(ASP Conf. Ser. 31, San Francisco, 1992), 169
\re
Poynting, J. H. 1903, Phil. Trans. Roy. Soc. London. Ser. A., 202, 525
\re
Pringle, J. E. 1996, MNRAS, 281, 357
\re
Rieke, G. H. 
1992, in Relationships Between Active Galactic Nuclei and
Starburst Galaxies, ed. Fillipenko, A. V.
(ASP Conf. Ser. 31, San Francisco, 1992), 61
\re
Robertson, H. P. 1937, MNRAS, 97, 423
\re
Rowan-Robinson, M. 1995, MNRAS, 272, 737
\re
Rowan-Robinson, M., \& Crawford, 1989, MNRAS, 238, 523
\re
Scoville, N. Z. 
1992, in Relationships Between Active Galactic Nuclei and
Starburst Galaxies, ed. Fillipenko, A. V.
(ASP Conf. Ser. 31, San Francisco, 1992), 159
\re
Scoville, N. Z., Sanders, D. B., Sargent, A. I., Soifer, B. T., Scott, S. L., \&
Lo, K. Y. 1986, ApJ, 311, L47
\re
Scoville, N. Z., Sargent, A. I., Sanders, D. B., \& Soifer, B. T. 
1991, ApJ, 366, L5
\re
Scoville, N. Z., Yun, M. S., \& Bryant, P. M. 1997, ApJ, 484, 702
\re
Shakura, N. I., \& Sunyaev, R. A. 1973, A\&A, 24, 337 
\re
Shapiro, P. S., \& Field, G. B. 1976, 205, 762
\re
Shlosman, I., Frank, J., \& Begelman, M. 1989, Nature, 338, 45
\re
Shore, S. N., \& White, R. L. 1982, ApJ, 256, 390
\re
Soifer, B. T., Sanders, D. B., Neugebauer, G., Danielson, G. E., Lonsdale, C. J.,
Madore, B. F., \& Persson, S. E. 1986, ApJ, 303, L41\
\re
Spitzer, L. Jr. 1962, in Physics of Fully Ionized Gases
(John Wiley \& Sons, Inc. 1962) Chapt. 5
\re
Spitzer, L. Jr. 1978, in Physical Processes in the Interstellar Medium
(John Wiley \& Sons, Inc. 1978) \S 9.3
\re
Storchi-Bergman, T., Wilson, A. S., \& Baldwin, J. A. 1996, ApJ, 460, 252
\re
Taniguchi, Y. 
1992, in Relationships Between Active Galactic Nuclei and
Starburst Galaxies, ed. Fillipenko, A. V.
(ASP Conf. Ser. 31, San Francisco, 1992), 357
\re
Taniguchi, Y. 1997, ApJ, 487, L17
\re
Terlevich, R. 
1992, in Relationships Between Active Galactic Nuclei and
Starburst Galaxies, ed. Fillipenko, A. V.
(ASP Conf. Ser. 31, San Francisco, 1992), 133
\re
Tomisaka, K., \& Ikeuchi, S. 1986, PASJ, 38, 697
\re
Tsuribe, T., Fukue, J., \& Umemura, M. 1994, PASJ, 46, 579
\re
Tsuribe, T., Umemura, M., \& Fukue, J. 1995, PASJ, 47, 73
\re
Tsuribe, T., \& Umemura, M. 1997, ApJ, 486, 48
\re
Umemura, M, \& Fukue, J. 1994, PASJ, 46, 567
\re
Umemura, M, Fukue, J., \& Mineshige, S. 1997, ApJ, 479, L97 (Paper I)
\re
Umemura, M, \& Ikeuchi, S. 1987, ApJ, 319, 601
\re
Umemura, M., Loeb, A., \& Turner, E. L. 1993, ApJ, 419, 459
\re
Wilson, A. S., Helfer, T. T., Haniff, C. A., \& Ward, M. J. 1991, ApJ, 381, 79
\re
Wada, K., \& Habe, A. 1995, MNRAS, 277, 433
\re
Yamada, T. 1994, ApJ, 423, L27
\re
Yi, I, Field, G. B., \& Blackman, E. G. 1994, ApJ, 432, L31
\re
Zeldovich, Ya, B., Ruzmaikin, A. A., \& Sokoloff, D. D. 1983, in
Magnetic Fields in Cosmology, The Fluid Mechanics of Astrophysics
and Geophysics, Vol. 3 (New York, Gordon and Breach Science
Publishers)

\newpage
\begin{figure}[hbtp]
\epsfxsize=16cm
\epsfbox{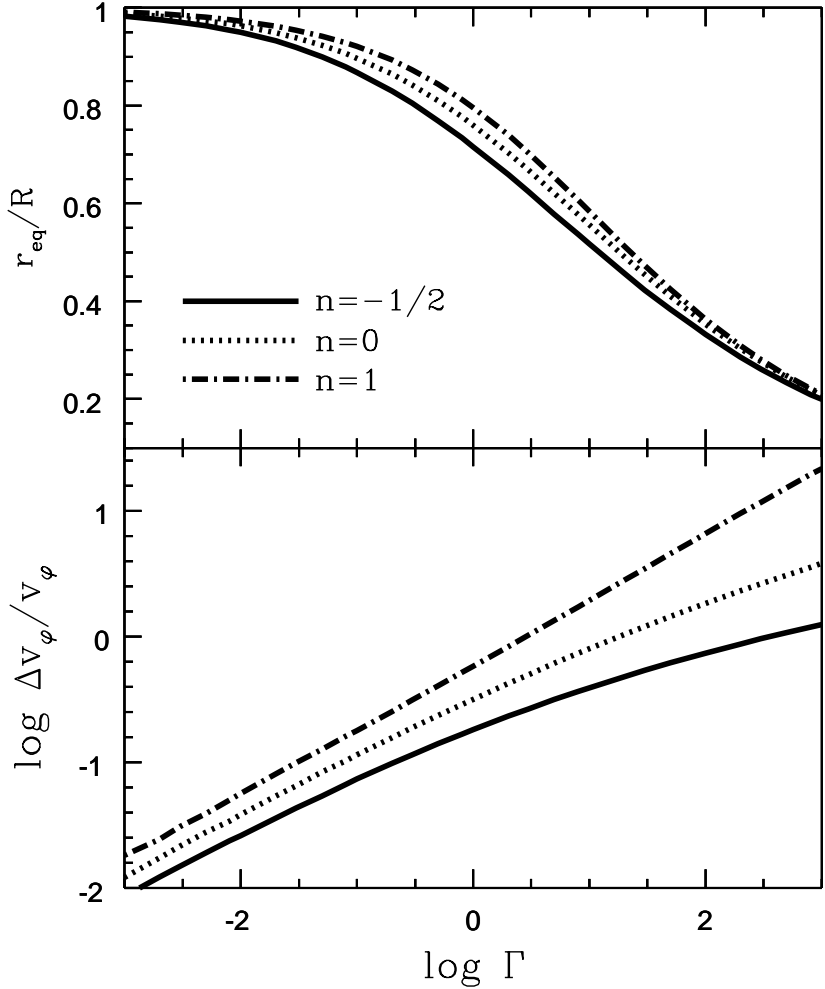}
\caption{The equilibrium radius, $r_{\rm eq}$, 
of the outer edge of a disk
after contraction by the radial radiation flux is plotted
against $\Gamma \equiv L_\ast/L_{\rm E}$ ({\it upper panel}), 
where $L_\ast$ is the bolometric luminosity of
a starburst ring and $L_{\rm E}$ is the Eddington luminosity corresponding to
the gravitational mass within the ring radius, $R$.
Here, three types of gravitational potentials are assumed, i.e.,
$n=-1/2$ (Keplerian rotation potential), $n=0$ (flat-rotation potential),
and $n=1$ (rigid-rotation potential).
Also, the fractional difference between 
an enhanced rotation velocity by the radiation flux and 
an original rotation velocity of gas disk,
$\Delta v_\varphi /v_\varphi$,
is shown at $r_{\rm eq}$ ({\it lower panel}).
}
\end{figure}

\newpage
\begin{figure}[hbtp]
\epsfxsize=16cm
\epsfbox{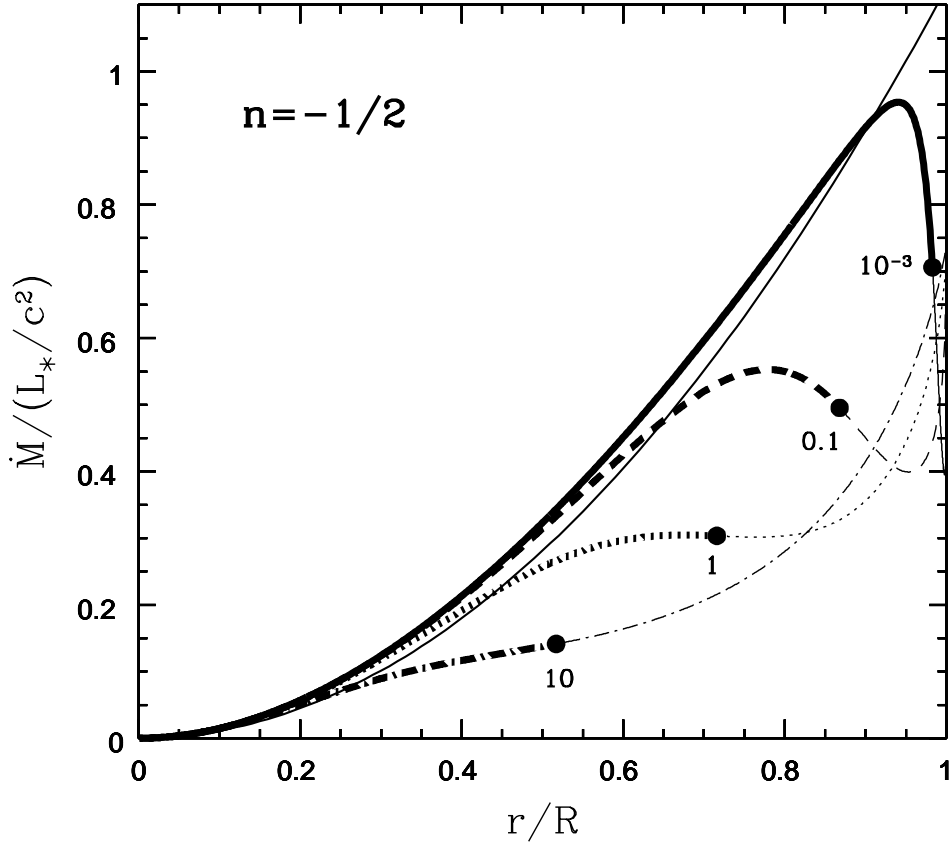}
\caption{(a) The mass accretion rate by a radiative avalanche 
is shown in units of $L_\ast/c^2$ as a function of radius 
normalized by the ring radius, $R$, for
$\Gamma \equiv L_\ast/L_{\rm E}=10$ ({\it dot-dashed curve}), 
1 ({\it dotted}), 0.1 ({\it dashed}), 
and $10^{-3}$ ({\it solid}).
Here, an optically thick disk is assumed in 
a gravitational potential of Keplerian rotation ($n=-1/2$).
Filled circles
represent the equilibrium radii of the outer-edge shown in Figure 1.
A thin solid curve is the previous estimation
with negligible effects of the radial flux:
$
\dot{M}/(L_\ast/c^2)=(3/8)(r/R)^2[(1-n)/(1+n)] 
$
(see Paper I). In the vicinity of the ring ($x \rightarrow 1$),
all solutions approach an asymptotic value of $\dot{M}=3L_\ast/4c^2$,
being independent of $\Gamma$ as well as $n$. 
}
\end{figure}

\setcounter{figure}{1}
\newpage
\begin{figure}[hbtp]
\epsfxsize=16cm
\epsfbox{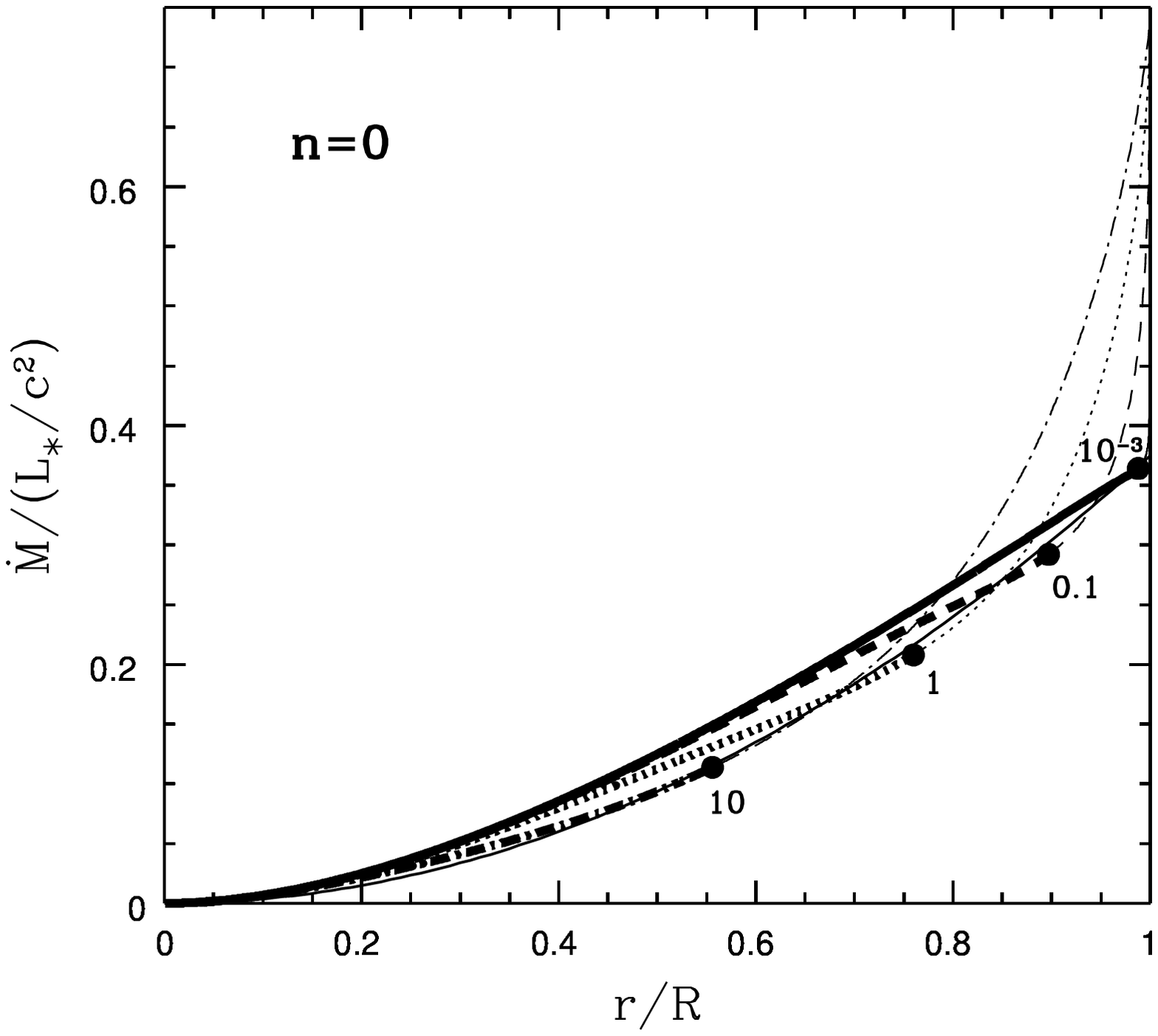}
\caption{(b) Same as Figure 2a, but for
$n=0$ (flat-rotation potential).}
\end{figure}

\setcounter{figure}{1}
\newpage
\begin{figure}[hbtp]
\epsfxsize=16cm
\epsfbox{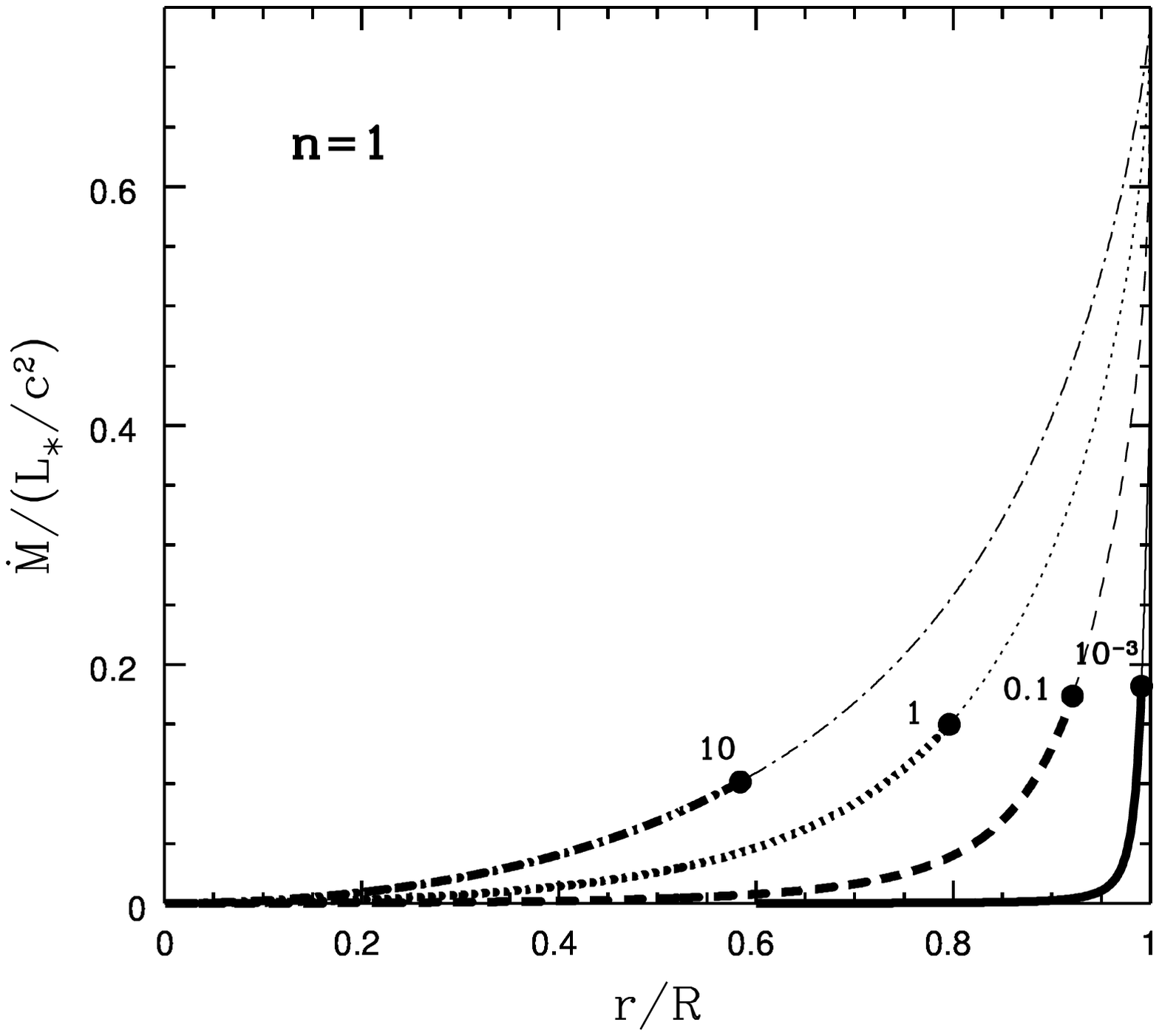}
\caption{(c) Same as Figure 2a, but for
$n=1$ (rigid-rotation potential).
Although the rigid rotation ($n=1$) results in no mass accretion
if a radial flux is negligibly small,
the radial flux for $\Gamma \geq 10^{-3}$
can generate an appreciable accretion.
}
\end{figure}

\newpage
\begin{figure}[hbtp]
\epsfxsize=16cm
\epsfbox{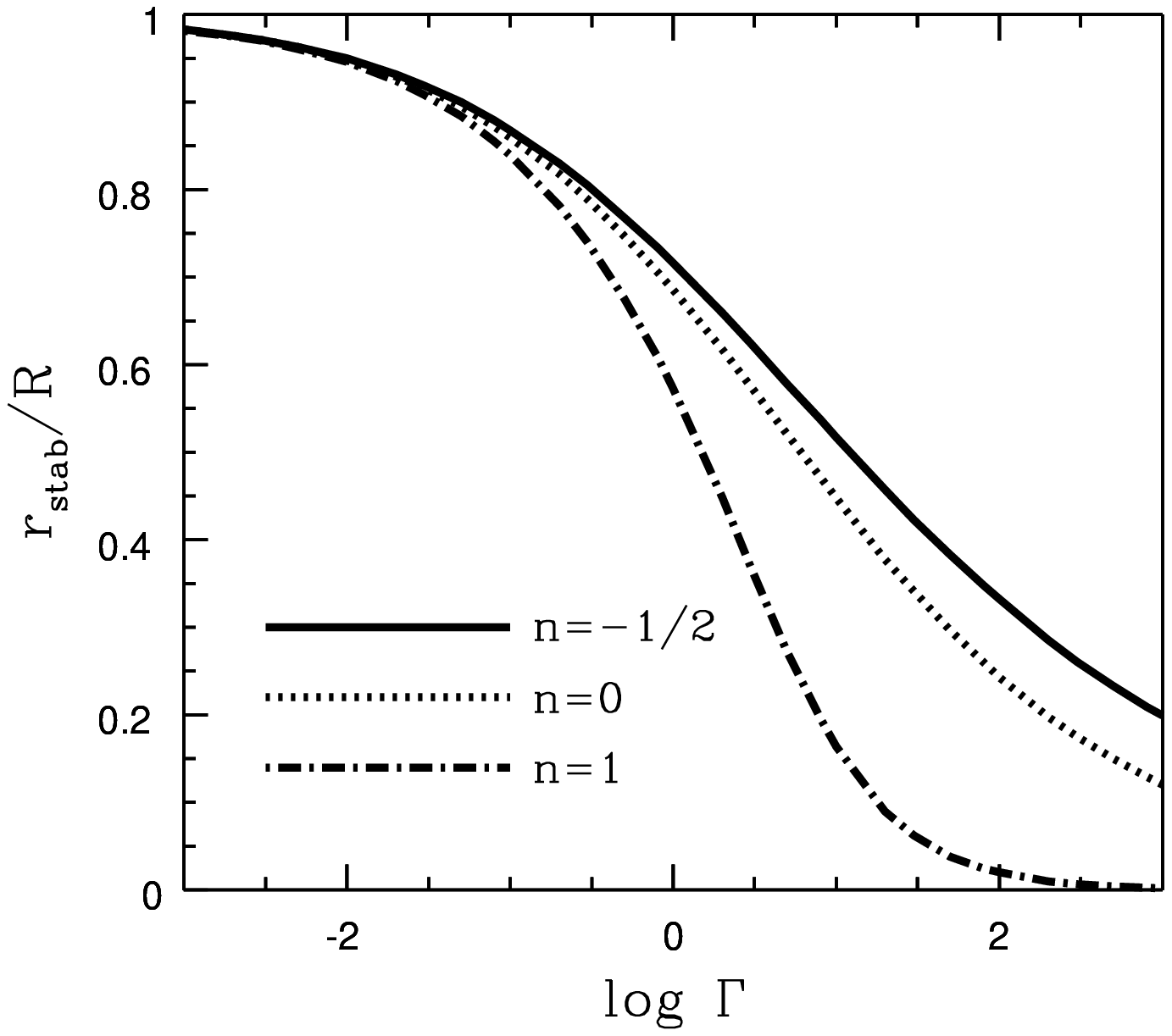}
\caption{The radius inside which the disk is vertically stable
is shown against $\Gamma \equiv L_\ast/L_{\rm E}$ for $n=-1/2$, 0, and 1.
Outside the radius, an optically thin disk or the outer edge of an optically
thick disk could be blown out by the radiation flux force.
}
\end{figure}

\newpage
\begin{figure}[hbtp]
\epsfxsize=16cm
\epsfbox{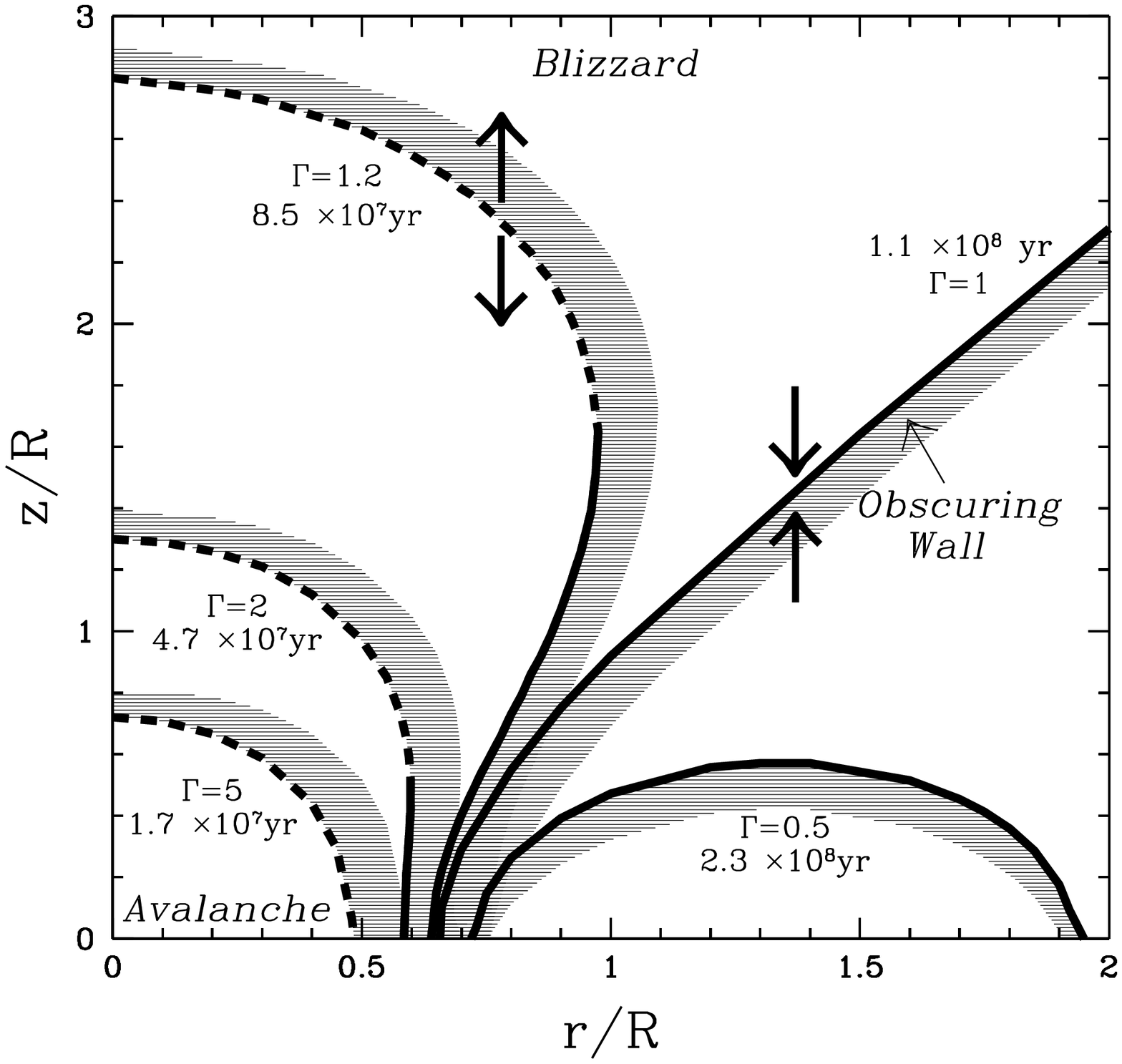}
\caption{Vertical radiation flux force is compared with the
vertical gravitational force for $n=-1/2$ (a Keplerian disk) 
in cylindrical coordinates $(r,~z)$. 
On the hatched sides for each value of $\Gamma\equiv L_\ast/L_{\rm E}$, 
the radiation force overwhelms the gravity. 
The dotted lines are corresponding to the unstable branches below which
gas is accelerated downward by the gravity, while above which
gas blows upward due to radiative force.
The solid lines are stable branches.
The attached ages are that of starburst regions corresponding to
each value of $\Gamma$.
}
\end{figure}

\end{document}